\documentclass[final,5p,times,twocolumn]{elsarticle}

\usepackage{amsmath}
\usepackage{amssymb}
\usepackage{bm}
\usepackage[english]{babel}

\journal{Physics Letters B}

\begin{document}

\begin{frontmatter}



\title{Critical Endpoint and Inverse Magnetic Catalysis for Finite Temperature and Density
Quark Matter in a Magnetic Background}

\author[label1]{M. Ruggieri\corref{cor1}}
\ead{marco.ruggieri@lns.infn.it}
\author[label1,label2]{L. Oliva}
\author[label1,label3]{P. Castorina}
\author[label4]{R. Gatto}
\author[label1,label2]{V. Greco}

\address[label1]{Department of Physics and Astronomy, University of Catania, Via S. Sofia 64, I-95125 Catania}
\address[label2]{INFN-Laboratori Nazionali del Sud, Via S. Sofia 62, I-95123 Catania, Italy}
\address[label3]{INFN-CT, Via S. Sofia 62, I-95123 Catania, Italy}
\address[label4]{Departement de Physique Theorique, Universite de Geneve, CH-1211 Geneve 4, Switzerland}
\cortext[cor1]{Corresponding author.}

\begin{abstract}
In this article we study chiral symmetry breaking for quark matter in a magnetic background, $\bm B$,
at finite temperature and quark chemical potential, $\mu$,
making use of the Ginzburg-Landau effective action formalism.
As a microscopic model to compute the effective action we use the renormalized quark-meson model.
Our main goal is to study the evolution of the critical endpoint, ${\cal CP}$, as a function of 
the magnetic field strength, and investigate on the realization of inverse magnetic catalysis
at finite chemical potential. We find that the phase transition
at zero chemical potential is always of the second order; 
for small and intermediate values of $\bm B$, ${\cal CP}$ moves towards 
small $\mu$, while for larger $\bm B$ it moves towards moderately larger values of $\mu$.
Our results are in agreement with the inverse magnetic catalysis scenario at finite
chemical potential and not too large values of the magnetic field, while at larger $\bm B$
direct magnetic catalysis sets in.

\end{abstract}

\begin{keyword}
Chiral transition with finite magnetic background and chemical potential
\sep Ginzburg-Landau effective action.

\PACS 21.65.Qr \sep 12.38.Mh \sep 12.38.Lg  


\end{keyword}

\end{frontmatter}


\section{Introduction}
Simulations of ultrarelativistic heavy ion collisions suggested the possibility that
huge magnetic fields are created during noncentral collisions~\cite{Kharzeev:2007jp,Skokov:2009qp,Voronyuk:2011jd}. 
The current estimate for the largest magnetic field produced is in the range
$eB/m_\pi^2 \approx 5 \div 15$, 
where $m_\pi$ corresponding to the pion mass in the vacuum
(to $eB = m_\pi^2$ corresponds $B \approx 10^{14}$ T).
These results triggered the study of the modifications a strong background field produces
on spontaneous chiral symmetry breaking of Quantum Chromodynamics (QCD) and on deconfinement,
both at zero and finite baryon density; for recent studies, as well as for some older results,
see~\cite{Kharzeev:2012ph,Bali:2011qj,Klevansky:1989vi,D'Elia:2011zu,Suganuma:1990nn,
Gusynin:1995nb,Frasca:2011zn,Klimenko:1990rh,Fraga:2008qn,
Agasian:2008tb,Bruckmann:2013oba,D'Elia:2010nq,Endrodi:2013cs,Fukushima:2010fe,Mizher:2010zb,Andersen:2012bq,
Skokov:2011ib,Fukushima:2012xw,Chernodub:2010qx,Buividovich:2008wf,Fraga:2012fs,Preis:2010cq,
Callebaut:2013ria,Andersen:2011ip,Burikham:2011ng,Filev:2010pm,Ferrari:2012yw,Gynther:2010ed,
Ferreira:2013oda,Costa:2013zca,Ferreira:2013tba,Kamikado:2013pya,Ruggieri:2013cya,
Colucci:2013zoa}.
The existence of strong fields in heavy ion collisions, combined to the excitation of QCD
sphalerons at high temperature, suggested the possibility of the 
Chiral Magnetic Effect~\cite{Kharzeev:2007jp,Fukushima:2008xe},
see~\cite{Kharzeev:2012ph} for reviews.  
Besides heavy ion collisions, even stronger magnetic fields might have been produced in the early universe
at the epoch of the electroweak phase transition, $t_{ew}$ \cite{Vachaspati:1991nm,Campanelli:2013mea}: a widely 
accepted value for the magnetic field at the transition
is $B(t_{ew}) \approx 10^{19}$ T, even if this value has rapidly decreased 
scaling as $a^{-2}$, where $a(t)$ denotes the scale factor of the expanding universe,
losing several order of magnitude at the QCD phase transition.   
Finally, relatively strong magnetic fields are relevant for magnetars,
$B \approx 10^{10}$ T~\cite{Duncan:1992hi}.
Therefore, there exist three physical contexts in which 
QCD in a strong magnetic background is worth to be studied.

In this letter, we address the problem of the chiral phase transition 
for quark matter at finite quark chemical potential, $\mu$, and nonzero magnetic field,
$\bm B$, focusing on the critical endpoint, ${\cal CP}$, of the phase diagram
where a second order and a first order transition lines meet each other,
and on the chiral phase transition at finite $\mu$.
In order to make quantitative predictions we build up a Ginzburg-Landau (GL)
effective potential for the chiral condensate as in \cite{Ruggieri:2013cya} with the inclusion
of a finite $\mu$, beside $T$ and $\bm B$ already considered in \cite{Ruggieri:2013cya}, .
Even if we restrict ourselves to the case of a homogeneous condensate,
the computation of the GL effective action has revealed a powerful tool to study
the transitions to inhomogeneous phases when these are of the second order \cite{Abuki:2011pf},
beside more general treatments relying on heat kernel expansion techniques \cite{Flachi:2010yz}.
In \cite{Abuki:2011pf} the coefficients of the GL potential are connected 
to those entering in the gradient expansion terms as well, which eventually trigger
inhomogeneous condensation.
Hence our calculations pave the way for an efficient computation of second order
transitions to inhomogeneous condensates  at finite $\bm B$ and $\mu$.
For the mapping of the phase diagram from the space of the GL coefficients to the
$T-\mu-\bm B$ space we need a microscopic model to compute the explicit
dependence of the GL coefficients on these variables. In this letter we make use
of the renormalized quark-meson model~\cite{Jungnickel:1995fp,Herbst:2010rf,Schaefer:2004en,Skokov:2010wb}. 
The advantage of this model is its renormalizability,
which allows to make quantitative predictions which are not affected by any ultraviolet scale.

In \cite{Ruggieri:2013cya} it was found that the critical point (${\cal CP}$)
at $\mu=0$ is not in the phase diagram;
hence it is of a certain interest to locate ${\cal CP}$ at finite $\mu$ and follow its
evolution as the strength of $\bm B$ is increased. Moreover we wish to study
the possible appearance of the phenomenon of inverse magnetic catalysis (IMC)
at finite $\mu$ \cite{Preis:2010cq,Andersen:2012bq}, that is te inhibition of spontaneous
chiral symmetry breaking by the magnetic field.
Our conclusions are that increasing the strength of $\bm B$ from zero
to small values results into
the evolution of ${\cal CP}$ towards smaller values of $\mu$, but this tendency
is reversed at strong $\bm B$. Hence within this model ${\cal CP}$ does not hit
the $\mu=0$ axis in the $T-\mu-\bm B$ space. Moreover we confirm the predicted
IMC scenario for small values of $\bm B$, at the same time offering a simple interpretation
of this phenomenon. On the other hand, for larger values of $eB$ we find
direct magnetic catalysis at finite $\mu$, that is, spontaneous
chiral symmetry breaking is favoured by the magnetic field.

\section{The model}
In this work we use the renormalized quark-meson model
as the microscopic model to compute the effective action at the chiral critical line.
The model and its renormalization have been already presented in detail
in a previous article \cite{Ruggieri:2013cya}, therefore here we remind only of the
relevant definitions and steps of renormalization which will be used here.

The lagrangian density of the model is given by
\begin{eqnarray}
{\cal L} &=& \bar q \left[iD_\mu\gamma^\mu - g(\sigma +
i\gamma_5\bm\tau\cdot\bm\pi)\right] q
\nonumber \\
&& + \frac{1}{2}\left(\partial_\mu\sigma\right)^2 +
\frac{1}{2}\left(\partial_\mu\bm\pi\right)^2 - U(\sigma,\bm\pi)~.
\label{eq:LD1}
\end{eqnarray}
In the above equation, $q$ corresponds to a quark field in the
fundamental representation of color group $SU(N_c)$ and flavor group
$SU(2)$; the covariant derivative, $D_\mu = \partial_\mu - Q_f e
A_\mu$, describes the coupling to the background magnetic field,
where $Q_f$ denotes the charge of the flavor $f$. Besides,
$\sigma$, $\bm\pi$ correspond to the scalar singlet and the
pseudo-scalar iso-triplet fields, respectively. The potential $U$
describes tree-level interactions among the meson fields,
\begin{equation}
U(\sigma,\bm\pi) = \frac{\lambda}{4}\left(\sigma^2
+\bm\pi^2-v^2\right)^2~, \label{eq:L2}
\end{equation}
which is invariant under chiral transformations. 

We restrict ourselves to the one-loop approximation as in \cite{Ruggieri:2013cya}. 
It has been shown in \cite{Skokov:2011ib,Kamikado:2013pya,Andersen:2012bq} that even including the
quantum fluctuations by means of the functional renormalization group
does not change the phase structure of the model.
In the integration process, the meson fields are fixed to their classical
expectation values, $\langle\bm\pi\rangle = 0$ and
$\langle\sigma\rangle \neq 0$. 
The physical value of $\langle\sigma\rangle$ will be
then determined by minimization of the thermodynamic potential.
This implies the replacement $g\sigma \rightarrow
g\langle\sigma\rangle$ in the quark action.
The field $\sigma$ carries the quantum numbers of the quark chiral condensate,
$\langle\bar q q\rangle$; hence, in the phase with $\langle\sigma\rangle\neq 0$,
chiral symmetry is spontaneously broken.

The one-loop thermodynamic potential associated to the interaction of fermions with
a magnetic background can be computed within the 
Leung-Ritus-Wang method~\cite{Ritus:1972ky}:
\begin{eqnarray}
\Omega_B &=& -N_c\sum_f \frac{|Q_f eB|}{2\pi}\sum_{n=0}^\infty\beta_n \nonumber \\
&&\times\int_{-\infty}^{+\infty}
\frac{dp_z}{2\pi}\left[E + 
T\sum_{\gamma = \pm 1}\log\left(1+e^{-\beta E_\gamma}\right)\right]~,
\label{eq:1}
\end{eqnarray}
where $n$ labels the Landau level, $E$ corresponds to the single particle excitation spectrum, 
\begin{equation}
E = \sqrt{p_z^2 + 2 |Q_f eB|n + m_q^2}~,
\label{Eq:Ene}
\end{equation}
and $m_q = g\langle\sigma\rangle$ is the constituent quark mass. The factor $\beta_n = 2 - \delta_{n0}$
counts the degeneracy of the $n^{th}$-Landau level. Finally $E_\gamma= \gamma\mu + E$.

The divergence in $\Omega_B$ is contained in the vacuum contribution. 
Since the model is renormalizable, we can treat this divergence by means of renormalization.
In order to prepare $\Omega_B$ for renormalization we add and subtract the contribution
at $\bm B = 0$, namely
\begin{equation}
\Omega_{0} = -2N_c N_f \int\frac{d^3p}{(2\pi)^3}\left[\omega 
+ T\sum_{\gamma = \pm 1}\log\left(1+e^{-\beta\omega_\gamma}\right)\right]~,
\label{eq:2}
\end{equation}
where $\omega = \sqrt{\bm p^2 + m_q^2}$ and $\omega_\gamma=\mu\gamma + \omega$. 
This procedure is convenient since it allows to collect
all the contributions due to the magnetic field into an addendum which is ultraviolet 
finite. Following the notation of \cite{Ruggieri:2013cya} we split 
$\Omega_0$ into the vacuum and the valence quark contributions, $\Omega_0=\Omega_0^0 + \Omega_0^T$ with
\begin{eqnarray}
\Omega_{0}^0 &=& -2N_c N_f \int\frac{d^3p}{(2\pi)^3}\omega~,\\
\Omega_{0}^T &=& -2N_c N_f T\int\frac{d^3p}{(2\pi)^3} 
\sum_{\gamma = \pm 1}\log\left(1+e^{-\beta\omega_\gamma}\right)~.
\end{eqnarray}
Hence we write
\begin{equation}
\Omega_B = \Omega_0 + \left(\Omega_B - \Omega_0\right) \equiv \Omega_0 + \delta\Omega~.
\label{eq:PP}
\end{equation}

In \cite{Frasca:2011zn,Ruggieri:2013cya} it has been proved
explicitly that $\delta\Omega$ is finite, modulo condensate independent 
terms, and it is not affected by renormalization. 
The condensate independent terms have been discussed in \cite{Endrodi:2013cs}, 
where it is pointed out that they affect the renormalization
procedure of electric charge and magnetic field, leaving however $eB$ invariant;
since $eB$ is the only quantity which couples to fermions in our model, we can safely
neglect this further renormalization.
Removing the UV divergences requires the addition of two counterterms to the thermodynamic potential,
\begin{equation}
\Omega^{c.t.} = \frac{\delta\lambda}{4}\frac{m_q^4}{g^4} + \frac{\delta v}{2}\frac{m_q^2}{g^2}~,
\label{eq:Vct}
\end{equation}
and the following renormalization conditions~\cite{Suganuma:1990nn,Frasca:2011zn}
\begin{equation}
\left.\frac{\partial (\Omega_0^0 + \Omega^{c.t.})}{\partial m_q}\right|_{m_q = g f_\pi}=
\left.\frac{\partial^2 (\Omega_0^0 + \Omega^{c.t.})}{\partial m_q^2}\right|_{m_q = g f_\pi}=0~,
\label{eq:r2}
\end{equation}
which amount to the requirement that the one-loop contribution
in the vacuum, namely $\Omega_0^0$, does not affect the expectation value
of the scalar field and the mass of the scalar meson. 
The total thermodynamic potential thus reads 
\begin{equation}
\Omega = \Omega_B + U + \Omega^{c.t.}~.
\label{eq:Otot}
\end{equation}

\section{Ginzburg-Landau expansion}
In this Section we present the novelty of our study. 
Our goal is to expand $\Omega$ in the Ginzburg-Landau (GL) form, in order to build up the
effective potential at the critical line for the order parameter:
\begin{equation}
\Omega = \frac{\alpha_2}{2}m_q^2 + \frac{\alpha_4}{4!}m_q^4
+ \frac{\alpha_6}{6!}m_q^6~.
\label{eq:i1}
\end{equation} 
This will be useful to compute the chiral critical line at finite $\mu$ and $\bm B$.
Given the thermodynamic potential in Eq. \eqref{eq:Otot}, the GL coefficients we need are 
obtained trivially as $\alpha_n = \partial^n\Omega/\partial m_q^n$
with derivative computed at $m_q=0$. The $\bm B$-dependence of the GL coefficients
comes only from $\delta\Omega$.
In the case $\mu=0$ we have found the analytical expressions for the GL 
coefficients \cite{Ruggieri:2013cya}; on the other hand, for $\mu\neq 0$ this has not been
possible because the presence of the quark chemical potential complicates the relevant
momentum integrals. Therefore in this work we rely on a numerical evaluation of the
GL coefficients. The kind of investigation we perform here is however still interesting:
in fact in \cite{Ruggieri:2013cya} it was found that at $\mu = 0$ the magnetic field does not
induce a first order phase transition; on the other hand it is known that at $\bm B=0$
and $\mu \neq 0$ a critical endpoint, ${\cal CP}$, appears for large enough values of $\mu$,
where the phase transition becomes of the first order. It is then of a certain interest
to study how ${\cal CP}$ evolves at finite $\mu$ and $\bm B$ to understand its fate
in the phase diagram as both $\bm B$ and $\mu$ are in the game. Moreover, the computation
of $\alpha_4$ will be crucial to explain the evolution of ${\cal CP}$ at finite
$\mu$, as we will discuss in the next Section.

Before presenting the numerical results for the general case $\bm B \neq 0$
and $\mu \neq 0$ it is useful to remind of a few particular results.
The critical temperature of a second order phase transition is obtained
as a solution of the equation $\alpha_2 = 0$; at $\mu=0$ and $\bm B=0$ this condition
implies \cite{Ruggieri:2013cya}
\begin{equation}
T_c^2 = \frac{6\lambda v^2}{g^2 N_c N_f}+ \frac{3g^2 f_\pi^2}{2\pi^2}~.
\label{eq:m2}
\end{equation}
For a numerical estimate we take the parameters of~\cite{Mizher:2010zb}, namely 
$\lambda=20$, $v=f_\pi$ and $g=3.3$; with this parameter set we find $T_c \approx 173$ MeV.
In the above equation no UV cutoff appears, as it would appear instead in NJL or NJL-like models,
see for example \cite{Frasca:2011bd}. In our calculation the UV cutoff dependence
has been removed by the renormalization and the only mass scale determining $T_c$
is $f_\pi$. Numerical estimate of $T_c$ in our case however is in agreement with 
the NJL calculations.
In the case of a very strong magnetic field the critical temperature 
can be obtained by looking at $\alpha_2$ in \cite{Ruggieri:2013cya}; we obtain
\begin{equation}
T_c^2 = 2Q_u^{Q_u}|Q_d|^{|Q_d|} |eB|~,
\label{eq:m3}
\end{equation}
which shows that $T_c \propto \sqrt{eB}$. 
This result is in contraddiction with recent lattice computations \cite{Bali:2011qj},
where it is found that in QCD $T_c$ decreases with the increase of $eB$. This
disagreement is most probably due to the lack of an appropriate description of the gluon sector
in the present model. It is however possible to improve the model itself in order
to describe the gluon backreaction to the magnetic field,
thus reproducing at least qualitatively the behaviour of the critical temperature,
see \cite{Chao:2013qpa,Ferreira:2013tba}. We will not consider here these complications,
leaving them to future studies.

\section{Phase diagram and critical endpoint}
\begin{figure}[t!]
\begin{center}
\includegraphics[width=7cm]{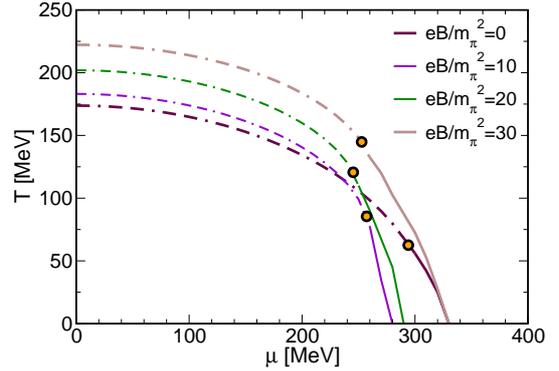}
\end{center}
\caption{\label{Fig:CL} Critical lines in the $T-\mu$ plane for several values of $eB$.
Dashed lines correspond to second order transitions, solid lines to first order transitions.
The critical endpoints for the different values of $eB$ are denoted by dots.}
\end{figure}

In Fig.~\ref{Fig:CL} we plot the phase diagram for spontaneous chiral symmetry breaking
in the $T-\mu$ plane. Dashed lines correspond to second order critical lines,
that are computed by solving the 
equation $\alpha_2(T_c,\mu,eB)=0$. As in previous model studies 
in which the vacuum contribution to the free energy is taken into account, the critical temperature
at zero and small $\mu$ is found to increase for increasing magnetic field strength.
As already said, this is not in agreement with recent lattice data which instead predict that
$T_c$ becomes smaller for increasing value of $eB$; it is clear that this 
discrepancy is not due to the lack of quantum fluctuations in the present model
calculations, see \cite{Skokov:2011ib,Kamikado:2013pya,Andersen:2012bq}.
Among the several possibilities suggested for the interpretation of this problem
\cite{Kojo:2012js,Chao:2013qpa,Ferreira:2013tba} the one closer to our work which does not
require the introduction of a Polyakov loop background is given in \cite{Chao:2013qpa} 
where an axial chemical potential, $\mu_5$, is added and its magnitude is assumed to be
an increasing function of $eB$, this dependence being inspired by previous works
which show that a large value of $eB$ increases
the fluctuations of chiral charge \cite{Buividovich:2008wf} 
and the sphaleron rate \cite{Basar:2012gh}. In fact a finite $\mu_5$
is found to decrease the temperature of chiral symmetry 
restoration  \cite{Fukushima:2010fe,Ruggieri:2011xc,Chernodub:2011fr}.
In the model at hand it is possible to
add the axial chemical potential, following the line of previous works within 
NJL as well as quark-meson model \cite{Fukushima:2010fe,Ruggieri:2011xc,Chernodub:2011fr}.
We will not consider this further complication here, leaving the inclusion of $\mu_5$
to a future project.
 
For completeness in Fig.~\ref{Fig:CL} we have also drawn the first order phase transition lines.
First order lines might be computed by the potential in Eq.~\eqref{eq:i1};
however the GL expansion is not expected to be quantitatively reliable at a 
first order line because the condensate might be still large
at the phase transition; therefore in order to compute those lines we have
used the full renormalized thermodynamic potential. Finally, the dots in the figure
denote the critical endpoints.

The critical lines depicted in Fig.~\ref{Fig:CL}
are in agreement with the scenario of inverse magnetic catalysis (IMC) at finite
$\mu$ \cite{Preis:2010cq,Andersen:2012bq}. More precisely for small and moderate
values of $eB$ and large enough $\mu$ the critical temperature decreases with 
increasing $eB$. For large values of $eB$ instead this IMC tendency seems to 
disappear and magnetic catalysis takes place.
Restricting the discussion to $T=0$
the IMC is evident for small fields since the critical value for chiral symmetry restoration,
$\mu_c$, decreases for increasing $eB$.
For larger values of the magnetic field instead we find that $\mu_c$ increases against $eB$. 
We can give handwaving arguments about why this phenomenon takes place within the model.
In this discussion it is useful to remind that  
$\mu_c$ is expected to be of the order of $m_q$.

In the model at hand, the restoration of chiral symmetry 
is due to the accomodation of valence quarks into single particle states, 
a process causing an increase of free energy that can be read from Eq.~\eqref{eq:1}, namely
\begin{equation}
\Delta\Omega_{vq}=\frac{N_c|eB|}{2\pi^2}\sum_f|Q_f|\sum_{n=0}^\infty\beta_n
\int_{0}^{+\infty}\!\!\! dp_z~ 
\theta(\mu-E)(E-\mu)~.
\label{eq:1a}
\end{equation}
The above contribution is finite and not affected
by renormalization. The $\theta-$function in Eq.~\eqref{eq:1a}
makes the integral nonvanishing only when the condition $\mu^2>m_q^2$
is satisfied. Moreover it implies that 
both the conditions $\mu^2 - m_q^2 > 2|eB|n$ and $\mu^2 - m_q^2 - 2|eB|n > p_z^2$
have to be satisfied. 
To measure energies from a common point we subtract from Eq.~\eqref{eq:1a} the analogous contribution
at $m_q=0$, since it corresponds to an irrelevant constant which does not modify
the value of the condensate and the transition point. Therefore we define
\begin{equation}
\delta\Omega_{vq}=\Delta\Omega_{vq} - \Delta\Omega_{vq}(m_q=0)~.
\end{equation}
Restricting to values of $\mu\approx m_q$ corresponding
to the regime where we expect a phase transition, the free energy gain corresponding to 
the accomodation of quarks into the phase space is
\begin{equation}
\delta\Omega_{vq}=\frac{N_c}{4\pi^2}|eB|m_q^2~,
\label{eq:1b}
\end{equation}
which can be derived from Eq.~\eqref{eq:1a} noticing that the restrictions
imposed by the $\theta-$function imply, for $\mu\approx m_q$, that only the 
lowest Landau level (LLL) gives a contribution to the sum if $eB$ is not too small. 
Equation~\eqref{eq:1b} shows that
the free energy gain  for accomodating valence quarks in the phase space is 
$\sim |eB|m_q^2$. On the other hand, if $eB$ is small enough then the renormalized
condensation free energy loss due to condensation in the magnetic field is \cite{Frasca:2011zn}
\begin{equation}
\delta\Omega_{c}=-\frac{N_c}{24\pi^2} (eB)^2\log\frac{m_q}{\lambda}~,
\label{eq:1c}
\end{equation}
where $\lambda$ plays the role of an infrared scale which does not affect the condensate.
The above equation corresponds to a negative contribution to the free energy
meaning that it favors the breaking of chiral symmetry 
because it lowers the value of $\Omega$.
It is easy to check that in this weak field limit, because of 
$m_q = g f_\pi + \delta m_q(eB)$ with $\delta m_q(eB)\sim (eB)^2/f_\pi^3$ \cite{Frasca:2011zn},
the further contributions to the condensation energy in the magnetic background
arising from $U$ and $\Omega_0^0$ are of the order $(eB)^4$ hence negligible
compared to Eq.~\eqref{eq:1c}.
Comparing Eqs.~\eqref{eq:1b} and~\eqref{eq:1c} we realize that the stabilization
in creating a condensate in the magnetic background is parametrically
smaller than the destabilization induced by the accomodation of valence quarks, 
therefore the net effect of the magnetic field will be to increase the
free energy of the condensed phase favouring the restoration of chiral symmetry.

On the other hand in the limit $eB \gg \mu^2$ the free energy loss due to condensation 
in the magnetic field is given by \cite{Frasca:2011zn}
\begin{equation}
\delta\Omega_{c}=-\frac{N_c}{8\pi^2} m_q^2|eB|\log\frac{|eB|}{m_q^2}~;
\label{eq:1d}
\end{equation}
the free energy gain $\delta\Omega_{vq}$
is still given by Eq.~\eqref{eq:1b}. In the 
strong field limit we realize a competition takes place between free energy loss
Eq.~\eqref{eq:1d} and gain Eq.~\eqref{eq:1c}, both being
of order $|eB|m_q^2$; moreover $\delta\Omega_{c}$ gets a logarithm enhancement 
for very large values of $eB$, which results eventually in 
lowering the free energy of the condensate phase enhancing chiral symmetry breaking. 
In this limit we expect catalysis of chiral symmetry breaking with $\mu_c^2$ proportional to $|eB|$,
which explains why we find that $\mu_c$ increases with $eB$ for large enough
values of $eB$. It is useful to notice that in the case we do not renormalize
the model and keep a finite value of the cutoff, $\Lambda$, then the logarithm
in Eq.~\eqref{eq:1d} is replaced by a function of $m_q/\Lambda$ as it can be proved
easily from Eq.~\eqref{eq:1} in LLL approximation; in this case
$\delta\Omega_{c}$ is still of the order of $|eB|m_q^2$ but it is not easy to predict
the fate of $\mu_c$ because the dependence of $m_q$ on $eB$ makes the comparison
of $\delta\Omega_{c}$ and $\delta\Omega_{vq}$ less transparent.

In Fig.~\ref{Fig:CL} the dots denote the critical endpoint, ${\cal CP}$,
in the $T-\mu$ plane for several values of $eB$. 
${\cal CP}$ is defined as the intersection of a second order
and a first order transition lines: for each value of $eB$ the ${\cal CP}$ coordinates are located
by solving the equations $\alpha_2(T,\mu,eB)=\alpha_4(T,\mu,eB)=0$.
The evolution of ${\cal CP}$ depicted in Fig.~\ref{Fig:CL} is quite peculiar since it shows
that increasing the value of $eB$ then ${\cal CP}$ does not hit the axis $\mu=0$;
rather it evolves towards large temperature and chemical potential.
The absence of ${\cal CP}$ at $\mu=0$ even for large magnetic fields
can be understood at the light of the results of \cite{Ruggieri:2013cya}:
at $\mu=0$ and very large $eB$ it has been found
\begin{equation}
\alpha_4\propto |eB|/T^2~,
\label{eq:TP}
\end{equation}
showing that the quartic coefficient of the GL expansion is always positive,
hence making the transition at $\mu=0$ a second order one for any value of $eB$.
The result in Eq.~\eqref{eq:TP} is obtained within the renormalized model; the use of
an explicit cutoff makes $\alpha_4$ negative at large enough $T$, thus turning the transition
to a first order and a critical point appears also at $\mu=0$.
If we use a fixed cutoff we expect thus that ${\cal CP}$ evolves towards the $\mu=0$ axis
for large enough $eB$. However we do not insist on this aspect because we are interested to
the phase structure of the renormalized model in which no explicit ultraviolet cutoff is present.
As a final comment we notice that the evolution of ${\cal CP}$ in Fig.~\ref{Fig:CL} 
is in agreement with an independent calculation making use of a model which takes into account
the Polyakov loop thermodynamics \cite{Costa:2013zca}.

\begin{figure}[t!]
\begin{center}
\includegraphics[width=7cm]{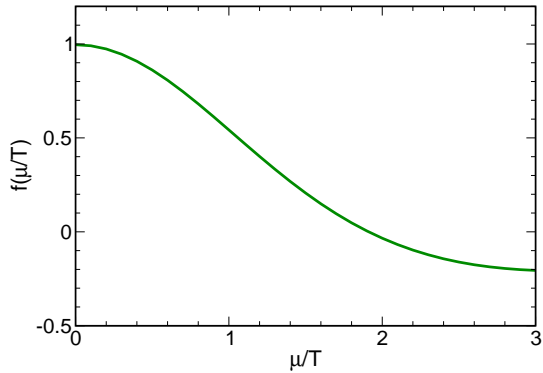}
\end{center}
\caption{\label{Fig:EX}Function $f$ entering in Eq.~\eqref{eq:AA4}.}
\end{figure}

The evolution of ${\cal CP}$ in the $T-\mu$ plane as a function of $eB$ in the model can be easily understood. 
For concreteness we refer to $eB/m_\pi^2 = 10$ and to
$eB/m_\pi^2 = 30$, because in between these two values of $eB$
the turning of ${\cal CP}$ evolution takes place. For the discussion the magnetic field dependent contribution
to $\alpha_4$, which we call $\delta\alpha_4$, have to be considered,
and $\alpha_4 = \alpha_4^0 + \delta\alpha_4$ with $\alpha_4^0 = \alpha_4(\bm B=0)$. 
We have checked that for the aforementioned values of 
magnetic field the higher Landau levels do not give a significant contribution to $\delta\alpha_4$ in the
critical region, therefore we do not include them
in the following discussion. In this case only the LLL contribution
to $\delta\alpha_4$ is necessary; a computation similar to that presented in \cite{Ruggieri:2013cya}
leads to the result
\begin{equation}
\delta\alpha_4 = \frac{3 N_c a_4}{\pi^2}\frac{|eB|}{T^2}f(\mu/T)~,
\label{eq:AA4}
\end{equation}
and $a_4\approx 0.11$. 
The function $f$ is shown in Fig.~\ref{Fig:EX}; in the $\mu\rightarrow 0$
limit Eq.~\eqref{eq:AA4} gives the result of  \cite{Ruggieri:2013cya}.  
For $eB/m_\pi^2 = 10$ the values of $\mu/T$ around ${\cal CP}$
are large enough to make $f$ negative, while $\alpha_4^0$ is positive. 
This means that LLL lowers the value of $\alpha_4$ favouring a first order phase transition.
This explains why ${\cal CP}$ moves towards smaller values of $\mu$.
On the other hand for $eB/m_\pi^2 = 30$ we find that $\alpha_4^0$
is suppressed compared to $\delta\alpha_4$ hence $\alpha_4\approx\delta\alpha_4$;  
moreover the values of $\mu/T$ in the critical region
are smaller because $T_c$ is enhanced by the magnetic field, eventually bringing
$f$ to be positive. As a result, in this case the LLL favours a second order phase transition,
thus pushing ${\cal CP}$ towards larger values of $\mu$. 

The evolution in Fig.~\ref{Fig:CL} is quite interesting because it shows that
increasing the strength of the magnetic field ${\cal CP}$ moves towards smaller
values of $\mu$ for moderate values of $eB$, then changing this tendency
for larger values of $eB$; this turning might suggest that the phase transition at $\mu=0$ 
becomes stiffer for moderate values of $eB$ then becomes softer, the stiffening and softening following
the evolution of ${\cal CP}$. However this is not the case and the phase transition at $\bm B=0$
becomes stiffer as $eB$ becomes larger. In fact one way to measure stiffness of the phase transition
is to compute $S\equiv |d m_q^2/dT|$ at $T=T_c$: from the potential \eqref{eq:i1} we get $m_q^2 = -6\alpha_2/\alpha_4$
(neglecting the $\alpha_6$ term, which can be done at a second order phase transition), which
for $T\lesssim T_c$ implies 
\begin{equation}
S = -\frac{6}{\alpha_4(T_c)}\left.\frac{d\alpha_2}{dT}\right|_{T=T_c}(T-T_c)~;
\end{equation} 
using the large field limit results of \cite{Frasca:2011zn}, namely
$\alpha_2\propto |eB|\log(|eB|/T^2)$ and $\alpha_4\propto |eB|/T^2$, we get
\begin{equation}
S \propto \frac{|eB|}{T_c}~,
\label{eq:SFm}
\end{equation}
which shows that the stiffness increases as $\sqrt{|eB|}$ since in the strong
field limit $T_c \propto \sqrt{|eB|}$. In the weak field limit one has to take into account
also the $\bm B-$independent contributions for $\alpha_2,\alpha_4$ but the correction to the stiffness
due to the magnetic field is still given by Eq.~\eqref{eq:SFm} in which, at the lowest
order, $T_c = T_c(\bm B=0)$, showing that in the weak field limit $S$ is enhanced as $|eB|$.
Summarizing, we find that $S$ increases with $eB$ both in the weak and in the strong field limit;
however the dependence on $eB$ is stronger in the weak field limit and weaker in the case of strong fields.

\section{Conclusions}
In this article we have studied the phase structure of hot quark matter in a magnetic background, $\bm B$,
at finite temperature, $T$, and quark chemical potential, $\mu$,
making use of the Ginzburg-Landau (GL) effective action formalism to compute the 
regions in the $T-\mu-eB$ space where chiral symmetry is spontaneously broken.
As a microscopic model to compute the GL coefficients we have used the renormalized quark-meson model.
The absence of an explicit ultraviolet cutoff permits a consistent calculation
even for large $\mu$ as well as for large $|eB|$. Apart from the work \cite{Ruggieri:2013cya}
which anticipates the formalism and some of the results we obtain here,
the renormalized quark-meson model has not been used for the study of the phase diagram of quark matter
at finite $\mu$ and $\bm B$. Therefore our study aims to fill this gap. 

The results obtained here for the critical temperature are in agreement with previous
studies based on different approaches. In particular we confirm the scenario of inverse
magnetic catalysis (IMC) at finite $\mu$ up to moderate values of $eB$,
in our calculations up to $eB\approx 10 m_\pi^2$; instead at large $eB$
magnetic catalysis appears. 
The IMC at small $eB$ is understood within this model
because the decrease of free energy due to condensation in magnetic field is parametrically
smaller than the increase of free energy necessary to accomodate valence quarks in the phase
space: in fact at small $eB$ for the former we have $\delta\Omega_{c}\sim-(eB)^2$
while for the latter $\delta\Omega_{vq}\sim m_q^2|eB|$ with $m_q\approx\mu$
at the phase transition. On the other hand at large $eB$
the renormalized decrease of free energy due to condensation is 
$\delta\Omega_{c}\sim-m_q^2|eB|\log(|eB|/m_q^2)$ and competes with $\delta\Omega_{vq}$,
eventually triggering magnetic catalysis thanks to the logarithm enhancement.

We have also computed the evolution of the critical endpoint ${\cal CP}$ in the
$T-\mu-eB$ space. We have found that for small and intermediate values of $eB$,
${\cal CP}$ moves towards smaller values of $\mu$; on the other hand
for large values of $eB$ the critical endpoint moves towards larger values
of $\mu$. We have explained this evolution in terms of the lowest Landau level contribution
to the coefficient $\alpha_4$ of the GL effective potential at finite $\mu$ and $T$, 
whose sign determines the order of the phase transition. 
This result agrees with the computation at $\mu=0$ of \cite{Ruggieri:2013cya}
where it was found that the $\alpha_4$ is always positive in the renormalized model
at $\mu=0$, thus favouring the scenario that at $\mu=0$ the phase transition
is of the second order also at large $eB$. 

There are several directions which are worth to be considered for continuing
the present work. Including an axial chemical potential following 
\cite{Ruggieri:2011xc,Chernodub:2011fr} is interesting in view of the possible role
this quantity has to induce inverse magnetic catalysis at $\mu=0$ \cite{Chao:2013qpa},
and study the interplay between $\mu$ and $\mu_5$ which was investigated
for the first time in \cite{Ruggieri:2011xc}. Moreover, the extension of the 
GL effective action formalism to study inhomogeneous 
phases \cite{Abuki:2011pf,Frolov:2010wn}
(see \cite{Anglani:2013gfu} for a review) is with no doubt fascinating.
Even more, the inclusion of the Polyakov loop thermodynamical
contribution to the effective potential is of a certain interest
because it might affect the GL effective action in a nontrivial way.
We plan to study these topics in our future projects.

\emph{Acknowledgements.} We acknowledge H.~Abuki, M. D'Elia and M.~Tachibana
for their careful reading of the manuscript and useful comments on the first version
of this letter.
V. G. acknowledges the ERC-STG funding under the QGPDyn
grant.



\end{document}